\documentclass[aps,
twocolumn,
jcp,
longbibliography,floatfix]{revtex4-2}

\usepackage[english]{babel}
\usepackage{amsmath,amsthm,latexsym,amssymb,amsfonts,epsfig}
\usepackage{bm}
\usepackage{dsfont}
\usepackage[left = 1.35cm, right = 1.35cm, top = 1.87cm, bottom = 1.87cm]{geometry}

\usepackage{graphicx}
\usepackage{color}

\definecolor{CarmineRed}{rgb}{1.0, 0.0, 0.22}
\definecolor{OliveGreen}{rgb}{0.1, 0.4, 0.1}

\newcommand{\R}{\bm{ r }}
\newcommand{\pder}[2]{\frac{\partial #1}{\partial #2}}
\newcommand{\pderline}[2]{\partial #1/\partial #2}

\newcommand{\Tinu}{\mathscr{ T }_{i\nu}}
\newcommand{\TinuP}{\mathscr{ T }_{i\nu+1}}
\newcommand{\TinuN}{\mathscr{ T }_{i\nu-1}}

\newcommand{\Kinukr}{K_{ i\nu } (|k|r)}
\newcommand{\invs}{\frac{ 1 }{ s }}
\newcommand{\phitil}{\widetilde{ \phi }}

\usepackage{setspace}
\emergencystretch=\maxdimen
\allowdisplaybreaks

\usepackage{mathtools}
\usepackage{mathrsfs}

\usepackage{type1ec} %

\usepackage{amsbsy}

\usepackage[colorlinks=true,citecolor=OliveGreen,linkcolor=OliveGreen,urlcolor=OliveGreen]{hyperref}

\newcommand{\ch}{\operatorname{ch}}
\newcommand{\sh}{\operatorname{sh}}
\newcommand{\cth}{\operatorname{cth}}

\newcommand{\csh}{\operatorname{csh}} 
\newcommand{\sch}{\operatorname{sch}} 

\newcommand{\sgn}{\operatorname{sgn}}

\begin{document}

\title{Spectral methods for wedge and corner flows: The Fourier–Kontorovich–Lebedev \\
integral transform}

\author{Abdallah Daddi-Moussa-Ider}
\email{abdallah.daddi-moussa-ider@open.ac.uk}
\affiliation{School of Mathematics and Statistics, The Open University, Walton Hall, Milton Keynes MK7 6AA, United Kingdom}

\begin{abstract}
    Understanding fluid flow in wedge-shaped geometries is essential for predicting hydrodynamic interactions in confined systems, such as microfluidic devices and near-corner transport phenomena. This article reviews analytical methods and techniques for addressing wedge problems in low-Reynolds-number hydrodynamics, focusing on solutions of the Stokes equations for a point force (Stokeslet) and a point torque (rotlet). 
    The formulation is based on the Papkovich–Neuber representation, which uses four harmonic functions to characterize the fluid flow.
    A concise overview of the Fourier–Kontorovich–Lebedev (FKL) transform method is provided, highlighting key properties and steps employed in deriving these solutions. This offers a versatile framework for predicting particle dynamics in wedge confinements and for designing microfluidic systems with corner geometries.
\end{abstract}

\maketitle

\section{Introduction}

Many physical phenomena are described by differential equations, such as the Laplace equation for potential fields or the Navier–Stokes equations for viscous fluid motion, whose solutions fully characterize the state and dynamics of the system under investigation. Solving these equations analytically is often highly challenging due to complex geometries or boundary conditions, which has motivated the development of powerful mathematical tools, including integral transform techniques, to obtain tractable representations of the solutions~\cite{erdelyi54, debnath2016integral, prudnikov1992integrals, abramowitz72, gradshteyn2014table, polyanin2008handbook}. Prime examples of these integral transforms include the Fourier and Laplace transforms, as well as the Hankel transform, particularly effective for problems in cylindrical geometries. By mapping differential equations into algebraic forms in the transformed space, these techniques greatly simplify the process of obtaining solutions.

Understanding boundary-value problems in wedge and corner geometries is essential across multiple areas of physics and engineering.
These problems are often tackled using the Kontorovich–Lebedev (KL) transform, sometimes in combination with other integral transforms such as Fourier or Laplace transforms.
This introduction begins by reviewing the origin and development of the KL transform, highlighting its mathematical foundations and early contributions. It then reviews key applications of the transform in the context of wave propagation, including electromagnetic and acoustic diffraction problems. Following this, the focus shifts to applications in linear elasticity, where the KL transform has been used to analyze stress and displacement fields in wedge-shaped solids. Finally, the discussion addresses the use of the KL transform in low-Reynolds-number flows near wedges and corners, which forms the central focus of this review article.

\subsection{The origin of the Kontorovich–Lebedev Transform}

The KL integral transform is a relatively less known yet powerful integral transform, particularly well suited for solving boundary-value problems in wedge-shaped or conical geometries. The method was first developed in the late 1930s by the Russian mathematicians Mikhail Iosifovich Kontorovich (1906--1987) and Nikolai Nikolaevich Lebedev (1911--1994), 
who were at that time based in Leningrad, now Saint Petersburg, Russia.
Using this approach, boundary-value problems in diffraction theory~\cite{kontorovich1938one, kontorovich1939method} and electrodynamics~\cite{kontorovich1939application} were solved by Kontorovich and Lebedev. The foundational mathematical theory of the integral transform was subsequently developed by Lebedev~\cite{lebedev1946, *lebedev1946french, lebedev1949, lebedev1974dual}.
A concise textbook presenting detailed solutions to applied mathematics problems—including applications of KL transforms—was coauthored by Lebedev together with Skal'skaia and Uflyand, first published in Russian and later translated and edited by Silverman~\cite[Ch.~6, p.~194]{lebedev1966problems}.

In contrast to more familiar transforms, the KL transform is classified as an index integral transform~\cite{wimp1964class, yakubovich1996index, polyanin2008handbook}. These form a class of integral operators in which the transformed function depends on an index parameter, typically linked to special functions~\cite{lebedev1972special}. 
Over recent years, numerous significant contributions to the foundations of the KL transform have been made by Yakubovich and his collaborators~\cite{yakubovich2003kontorovich, yakubovich2008progress, yakubovich2012index, rodrigues2013convolution, loureiro2013central, yakubovich2024index}. They advanced the theory of the KL transform by developing related integral and convolution operators and establishing connections with other integral transforms.
Together, these developments established the KL transform as a versatile and rigorous tool for tackling complex boundary-value problems in mathematical physics. For a compact overview of this outstanding integral transform, including tables of transforms for common special functions, see the classic textbook by Erd{\'e}lyi \textit{et~al.}~\cite[Ch.~XII, p.~173]{erdelyi54}.

\subsection{Applications in wave propagation}

Over the past few decades, the KL transform has been extensively applied to a wide range of physical problems, particularly those involving diffraction and scattering in electromagnetism and acoustics. These systems are typically governed by the wave equation, or equivalently by the Helmholtz equation in the time-harmonic regime, which arises from Maxwell’s equations in electromagnetism and from the linear acoustic wave equation for pressure fields~\cite{landau1975classical, landau1984electrodynamics}. In this context, a Laplace or Fourier transform allows the separation of variables in time and, if relevant, along the axial direction. When combined with the KL transform in the radial coordinate, the original multidimensional problem reduces to an ordinary differential equation in the polar angle. This approach is particularly effective for linear, time-dependent, or spatially translationally invariant problems, where transforms exploit the problem’s symmetries and boundary conditions.

Beyond the pioneering works of Lebedev~\cite{kontorovich1938one, kontorovich1939method, kontorovich1939application}, an early application of the KL transform was presented by Lowndes~\cite{lowndes1959application} to determine the electromagnetic field generated by a line current near a perfectly conducting wedge. Subsequently, the technique was applied by Jones to a wide range of electromagnetic and acoustic wave problems, as documented in his seminal textbooks~\cite{jones1964theory, jones1986acoustic}. Thereafter, the diffraction of elastic waves by a wedge was investigated by Forristall~\cite{forristall1970elastodynamics}, thereby elucidating the seismic response of topographic irregularities. The field was revitalized in the 1990s through the contribution of Davis~\cite{davis1996two}, in which acoustic diffraction near a penetrable wedge was analyzed. The KL transform was applied by Scharstein and Davis to extend the Biot–Tolstoy theory~\cite{biot1957formulation} to wedges with density contrast~\cite{davis1997complete} as well as to isorefractive wedges~\cite{scharstein1998time}. In a related development, the diffraction of electromagnetic waves by penetrable wedges was investigated by Rawlins~\cite{rawlins1999diffraction}.

The field has experienced substantial growth in the 21st century, as reflected by the increasing number of publications and research activity. Solutions for a diaphanous wedge were provided by Knockaert \textit{et~al.}~\cite{knockaert2002diaphanous}, while diffraction by circular cones was examined by Antipov~\cite{antipov2002diffraction}. 
Later, the electromagnetic field generated by a line source near a dielectric wedge was studied by Salem \textit{et~al.}~\cite{salem2006electromagnetic}. 
The fundamental solution for a point-source potential in a wedge composed of two homogeneous media was derived by Scharstein~\cite{scharstein2004green}. 
Further applications of the KL transform include scattering from a grooved conducting wedge, investigated by Hwang~\cite{hwang2009scattering}, and scattering from a slotted conducting wedge, studied by Kim \textit{et~al.}~\cite{kim2009electromagnetic}.
A modified KL transform was subsequently developed by Shanin and Valyaev~\cite{shanin2011modified}.
The transform was applied by Lyalinov to various acoustic diffraction problems involving wedges or cones~\cite{lyalinov1999diffraction, lyalinov2007acoustic, lyalinov2009far, lyalinov2016integral, lyalinov2022eigenoscillations}. A comprehensive overview of techniques for electromagnetic scattering, covering a variety of transforms, was provided by Eom~\cite{eom2014integral}.

\subsection{Applications in linear elasticity}

Unlike its extensive use in electromagnetic and acoustic diffraction, the KL transform has seen comparatively limited application in linear elasticity~\cite{landau1986theory}. 
The integral transform has been used to derive analytical solutions of the Navier–Cauchy equations for wedge-like geometries, providing a powerful tool for addressing stress and displacement fields in solids with corners. Early applications by Kassir and collaborators~\cite{kassir1973application, kassir1975thermal} focused on solving crack problems using Papkovich–Neuber representations, yielding the stress distribution within the solid. 
To understand crack propagation and stress concentration in layered wedge-shaped structures, a periodic crack problem in an elastic layered wedge was analyzed by Pozharskii \textit{et al.}~\cite{pozharskii2020periodic}. In a related study, strip and elliptical cuts located in the middle of an elastic wedge were investigated by Aleksandrov and Pozharskii~\cite{aleksandrov2009problems}.

The Green’s function for the displacement on the surface of an incompressible wedge subjected to a point force was derived by Hanson and coworkers~\cite{hanson1991analysis, hanson1994stress}. 
These studies were motivated by the need to examine edge effects in rail–wheel systems, modeling the rail head as a three-dimensional elastic wedge to analyze contact stress and geometry.
The fundamental solution for a three-dimensional wedge was used by Aleksandrov and Pozharskii~\cite{aleksandrov2002problem} to study problems involving a thin, rigid, elliptic inclusion within a wedge and later generalized to composite wedges~\cite{aleksandrov2011three}. 
More recently, the Green's function for a linearly elastic, homogeneous, isotropic material in a wedge-shaped geometry was derived by Daddi-Moussa-Ider \textit{et al.} for forces both parallel~\cite{daddi2025proc} and perpendicular~\cite{daddi2025jelasticity} to the wedge edge, taking into account combinations of no-slip and free-slip boundary conditions.

\subsection{Applications in low-Reynolds-number flows}
The study of low-Reynolds-number flows near three-dimensional corners has provided fundamental insights into viscous fluid behavior in corners.
Stokes flows in wedge geometries were first studied by Hasimoto and collaborators, who analyzed a sequence of wedge geometries, starting with a right-angled wedge~\cite{sano76}, followed by obtuse wedges~\cite{sano1977slow, sano1978effect, hasimoto80, sano1977slow_thesis}, and semi-infinite planes~\cite{hasimoto1983effect}.
The case of salient wedges was later investigated by Kim~\cite{kim1983effect}. In these problems, the solution is obtained using a Fourier transform along the axial direction and a KL transform along the radial direction. These studies enabled the calculation of hydrodynamic mobility functions of colloidal particles in the far-field limit.
Remarkably, it has been shown that when the wedge opening angle falls below a threshold value, an infinite sequence of eddy rings forms around the direction of the point force~\cite{sano1980three}.
The Stokes flow in a semi-infinite wedge was later investigated by Shankar using the method of vector eigenfunctions~\cite{shankar2000stokes}.

The leading-order flow behavior both near the corner and in the far field for three-dimensional Stokes flow was characterized by Dauparas and Lauga~\cite{dauparas2018leading, dauparas2018stokes} through the combination of asymptotic expansions with complex analysis.
Using the method of images, the Green’s function for free-slip boundary conditions on wedges with commensurate opening angles was obtained by Sprenger and Menzel~\cite{sprenger2023microswimming}.
More recently, the flow induced by a locally applied point torque (rotlet) under no-slip boundary conditions on wedge surfaces was derived by Daddi-Moussa-Ider \textit{et~al.}~\cite{daddi2026hydrodynamic}.
The study complemented earlier results by Sano and Hasimoto~\cite{sano1978effect} by providing the elements of the rotational mobility of colloidal particles in wedge confinement.
In a similar vein, the Green’s function for a point source near a wedge was derived by Daddi-Moussa-Ider and Golestanian~\cite{daddi2026toward}, enabling the analysis of self-diffusiophoretic propulsion of a catalytically active particle close to a corner.

Although the KL transform was historically developed in wave physics, the present review concentrates on its role in viscous flows near wedges and corners. We revisit key techniques introduced in earlier works and summarize the solution methodology as adapted to incompressible Stokes flow. 
We focus on the steady Stokes equations, allowing the analysis to concentrate on spatial variations of the flow within wedge geometries.
Since our approach combines Fourier transforms along the axial direction with the KL transform, we refer to this as the Fourier–Kontorovich–Lebedev (FKL) transform, representing a synthesis of both techniques.

Our discussion is restricted to systems in which the fluid motion is driven by either a point force or a point torque, and we focus on the physically relevant case of no-slip boundary conditions imposed on the wedge surfaces. 
Other types of boundary conditions, such as the free-slip condition, can also be analyzed within the same methodological framework presented in this article~\cite{daddi2025proc, daddi2025jelasticity}.
The purpose of this review is to progressively develop the theoretical approach underlying the FKL transform, clarify its practical implementation for wedge geometries, and consolidate existing results to provide a coherent foundation for future advances in corner and wedge hydrodynamics.
This review serves as an introductory read for those wishing to explore the modeling of flow problems near wedges and corners, progressively presenting the underlying mathematical framework and methods.

The remainder of the article is organized as follows. In Sec.~\ref{sec:mcgovern_eqns}, we present the equations governing fluid motion and outline the framework used to obtain solutions based on the Papkovich–Neuber representation. Section~\ref{sec:fkl} introduces the FKL transform, first by reviewing the relevant properties, and then by representing the solution form in FKL space. In Sec.~\ref{sec:solution}, we summarize previously known solutions for a point force and point torque acting in a wedge-shaped geometry, oriented either parallel or perpendicular to the edge of the wedge. Finally, Sec.~\ref{sec:discussion} provides a discussion and concluding remarks.

\section{Governing equations}
\label{sec:mcgovern_eqns}

\begin{figure}
    \centering
    \includegraphics[width=\linewidth]{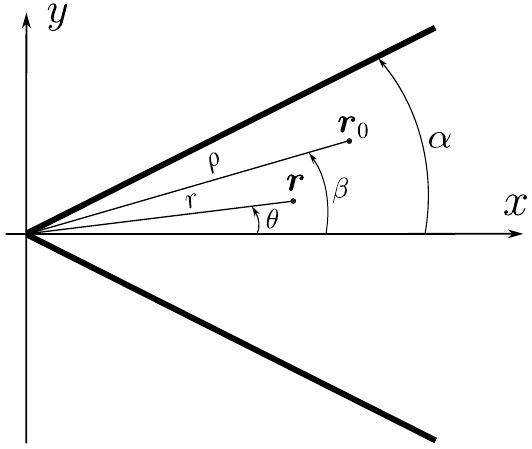}
    \caption{Schematic of the system representing a fluid confined within a wedge-shaped domain bounded by two planar walls. The geometry is described in cylindrical coordinates $\R = (r, \theta, z)$, with $r$ the radial distance from the wedge apex, $\theta$ the polar angle from the centerline between the walls, and $z$ the axial direction along the wedge edge. No-slip boundary conditions (zero velocity) are applied at $\theta = \pm \alpha$, and a flow singularity, either a point force or a point torque, is located at $\R_0 = (\rho, \beta, 0)$}
    \label{fig:illustration}
\end{figure}

The system under investigation is three dimensional and consists of a viscous, incompressible fluid confined within a wedge-shaped domain, bounded by two walls located at $\theta = \pm \alpha$.
The upper and lower walls of the wedge are characterized by the normal unit vectors $\left(\sin\alpha, -\cos\alpha, 0\right)^\top$ and $\left(\sin\alpha, \cos\alpha, 0\right)^\top$, respectively, both pointing toward the fluid domain.
The intersection of the two walls forms a corner with a wedge semi-opening angle $\alpha \in (0, \pi/2]$; see Fig.~\ref{fig:illustration} for a graphical illustration of the system setup.
The geometry is described in cylindrical coordinates $(r, \theta, z)$, where $r$ is the radial distance from the wedge apex, $\theta$ is the polar angle measured from the centerline between the walls, and $z$ is the axial direction along the wedge edge, which is invariant under translation. 
For $\alpha = \pi/4$, the wedge forms a right angle at the apex, whereas the planar wall limit is recovered for $\alpha=\pi/2$.
No-slip boundary conditions, corresponding to zero fluid velocity, are imposed at $\theta = \pm \alpha$.

We refer to a \textit{singularity} as an idealized, localized force or torque acting at a single point in the fluid, represented mathematically by a Dirac delta function. A \textit{point force singularity}, also known as a Stokeslet, corresponds to a concentrated force applied at a point, generating a flow field that decays away from that location but is formally infinite at the point itself. Similarly, a \textit{point torque singularity}, also known as a rotlet, represents a localized torque acting at a point, inducing a rotational flow field that is likewise singular at the point of application. Such singularities constitute fundamental solutions of the Stokes equations and serve as building blocks for more complex flow configurations. In the present work, we assume that the flow singularity, whether a point force or a point torque, is located in cylindrical coordinates at $(r, \theta, z) = (\rho, \beta, 0)$.

At low Reynolds numbers, where the characteristic length and velocity scales of the flow are small, inertial effects become negligible compared with viscous forces. Under these conditions, the fluid motion is governed by the Stokes equations~\cite{kim05},
\begin{equation}
 \label{eq:Stokes}
    -\boldsymbol{\nabla} p(\R) + \eta \boldsymbol{\nabla}^2 \bm{v} (\R) + \bm{f}(\R) = \bm{0} \, , \quad
    \boldsymbol{\nabla} \cdot \bm{v} (\R) = 0 \, ,
\end{equation}
where $\bm{v}$ and $p$ denote the fluid velocity and pressure fields, respectively, and $\eta$ is the dynamic viscosity.
Here, $\bm{f}$ denotes the force density. For a point force of strength~$\bm{F}$ applied at $\bm{r}_0$, it is given by $\bm{f}(\R) = \bm{F} \delta(\R-\bm{r}_0)$, commonly referred to as a \textit{Stokeslet}. For a point torque of strength~$\bm{T}$, it is expressed as $\bm{f}(\R) = \tfrac{1}{2} \, \bm{T} \times \boldsymbol{\nabla} \delta(\R-\bm{r}_0)$, known as a \textit{rotlet}, which generates a rotational or swirling flow.
Our aim is to determine the flow generated by a point force or torque singularity within a wedge geometry.

The general solution of Eq.~\eqref{eq:Stokes} can be expressed using the Papkovich–Neuber representation as~\cite{Papkovich1932,Neuber1934, tran1982general}
\begin{equation}
    \bm{v} (\R) = \bm{\nabla} \bigl( \R \cdot \boldsymbol{\Phi} (\R) + \rho \, \Phi_w (\R) \bigr) - 2 \, \boldsymbol{\Phi} (\R) \, , 
    \label{eq:Papkovich_Neuber}
\end{equation}
where $\boldsymbol{\Phi} (\bm{r})$ consists of the Cartesian components $\Phi_x$, $\Phi_y$, and $\Phi_z$. 
This representation is sometimes attributed to Imai~\cite{imai1973ryutai} in the Japanese literature.
In the following, we omit the explicit dependence of $\bm{v}$ and $\boldsymbol{\Phi}$ on $\R$. Each component $\Phi_j$ is a harmonic function satisfying Laplace’s equation,
\begin{equation}
    \boldsymbol{\nabla}^2 \Phi_j = 0 \, , \quad \text{for}\,\, j \in \{x, y, z, w\} \, .
\end{equation}

In cylindrical coordinates, the velocity field is represented by its radial, azimuthal, and axial components, $v_r$, $v_\theta$, and $v_z$, respectively. Applying the standard Cartesian-to-cylindrical transformations to Eq.~\eqref{eq:Papkovich_Neuber}, the velocity components in this coordinate system become
\begin{equation}
   v_r = \pder{\Pi}{r} - 2\Phi_r \,, \,\,
v_\theta = \frac{1}{r} \pder{\Pi}{\theta} - 2\Phi_\theta \,, \,\,
v_z = \pder{\Pi}{z} - 2\Phi_z \, .
   \label{eq:velocity_cylindrical}
\end{equation}
Here, 
\begin{equation}
    \Pi = r\Phi_r + z\Phi_z+\rho \Phi_w \, ,
    \label{eq:Pi}
\end{equation}
and
\begin{subequations}
    \label{eq:Phi_r_Phi_The}
    \begin{align}
    \Phi_r &= \Phi_x\cos\theta + \Phi_y\sin\theta \, , \\
    \Phi_\theta &= \Phi_y\cos\theta -\Phi_x \sin\theta \, .
\end{align}
\end{subequations}
The functions $\Phi_r$ and $\Phi_\theta$ are not harmonic, a point that should be emphasized to avoid confusion in the subsequent developments.

In Stokes flow, the solution is commonly represented in terms of three independent harmonic functions, known as Lamb's general solution~\cite{happel12}.
These are the harmonic velocity potential, the harmonic stream function, and the harmonic pressure field. Since the Papkovich–Neuber representation involves four harmonic functions, one of them is necessarily linearly dependent on the others. However, regardless of the specific choice made, the same physical solution is represented by this formulation.

The general solution of the flow equations in a wedge-shaped geometry for the four harmonic functions is expressed as
\begin{equation}
    \label{eq:solution_decompo}
    \Phi_j = q_{l} \left( \phi_j^\infty + \phi_j \right) \, , \quad \text{for}\,\, j \in \{x, y, z, w\} \, ,
\end{equation}
wherein $\phi_j^\infty$ denotes the harmonic functions associated with the solution in an unbounded fluid, referred to here as the ``free-space'' solution, i.e., the solution far away from all boundaries. The auxiliary fields $\phi_j$ are four unknown harmonic functions, representing the complementary solution.
These are required to satisfy the boundary conditions on the wedge surfaces.
In addition, we define
\begin{equation}
    q_{l} = \frac{F_{l}}{8\pi\eta} \, ,
\end{equation}
for a point force of magnitude $F_{l}$, and
\begin{equation}
    q_{l} = \frac{T_{l}}{16\pi\eta \rho} \, ,
\end{equation}
for a point torque of magnitude $T_{l}$.
Here, ${l} \in \{\parallel, \perp\}$, where $\parallel$ and $\perp$ indicate that the point force or torque is directed parallel or perpendicular, respectively, to the edge of the wedge along the $z$-direction.

Imposing no-slip boundary conditions on the wedge surfaces, Eqs.~\eqref{eq:velocity_cylindrical} suggest that a natural choice for the boundary conditions at $\theta = \pm \alpha$ is
\begin{equation}
    \label{eq:NS_BCs}
    \Phi_r = 0 \, , \quad
    \Phi_z = 0 \, , \quad
    \Phi_w = 0 \, , \quad
    \frac{1}{r}
    \pder{\Pi}{\theta}  - 2\Phi_\theta = 0 \, ,
\end{equation}
imposed at $\theta = \pm\alpha$.
Accordingly, we require $\Pi = 0$ at $\theta = \pm \alpha$, ensuring that $\pderline{\Pi}{r} = \pderline{\Pi}{z} = 0$ at the boundaries.
This is merely a choice and is not unique, as discussed above, since there is flexibility in selecting the fourth dependent harmonic function. Because this choice greatly simplifies the equations for the radial and axial components, setting $\Pi=0$ at the boundary is the standard approach when studying wedge problems using the Papkovich–Neuber representation.

The free-space velocity solution takes different forms depending on whether a point force or a point torque acts on the fluid. For a point force, the velocity in an unbounded viscous fluid is given by the Oseen tensor~\cite{oseen28, happel12}
\begin{equation}
    \bm{v}^\infty = 
    \frac{1}{8\pi \eta s}
    \left( \bm{I} + \hat{\bm{s}} \hat{\bm{s}} \right) \cdot \bm{F} \, ,
    \label{eq:free_space_force}
\end{equation}
where $\bm{I}$ denotes the identity tensor, $\bm{s} = \bm{r} - \bm{r}_0$ is the relative position vector from the singularity to the observation point, $s = |\bm{s}|$ is the distance between them, and $\hat{\bm{s}} = \bm{s}/s$ is the associated unit vector.
This velocity field decays in the far field proportionally to the inverse of the distance.

For a point torque, the velocity field is given by
\begin{equation}
    \bm{v}^\infty  = \frac{\bm{T} \times \hat{\bm{s}}}{8\pi\eta s^2}  \, ,
    \label{eq:free_space_torque}
\end{equation}
which decays in the far field proportionally to the inverse square of the distance.


\section{Fourier–Kontorovich–Lebedev transform}
\label{sec:fkl}

\subsection{General overview}

The FKL transform is a two-step integral transform used to simplify the analysis of problems involving wedge geometries.
In this approach, the Fourier transform is first applied along the axial coordinate, converting the dependence on this variable into a spectral representation. Subsequently, the KL transform is applied to the radial coordinate of the function, effectively mapping the radial dependence onto a continuous spectral parameter. This combined application of the Fourier and KL transforms reduces the original partial differential equations governing the system to a set of ordinary differential equations in the polar angle, greatly facilitating both analytical and numerical treatments.

In this review article, the hyperbolic sine, cosine, tangent, and cotangent functions are denoted by sh, ch, th, and cth respectively. Additionally, sch and csh represent the hyperbolic secant and cosecant hyperbolic functions, respectively, defined as $\sch \zeta =1/\ch \zeta$ and $\csh \zeta = 1/\sh \zeta$.

We define the forward Fourier transform of a function $f(r,z)$ with respect to $z$ as
\begin{equation}
    \label{eq:Fourier_forward}
    \hat{f}(r, k) 
    := \mathscr{F} \left\{ f \right\}
    = 
    \int_{ -\infty }^{\infty} 
    f(r,z) \, e^{ikz} \, \mathrm{d} z \, , 
\end{equation}
adopting the convention of a positive exponent for the forward transform. The KL transform with respect to $r$ is then applied to the Fourier-transformed function as
\begin{equation}
    \label{eq:KL_forward}
    \widetilde{f}(\nu, k) 
    := \mathscr{K}_{i\nu} \left\{ \hat{f} \right\}
    = 
    \int_{0}^{\infty} \hat{f}(r, k) \, \Kinukr \, r^{-1} \, \mathrm{d} r  \, ,
\end{equation}
with $\Kinukr$ denoting the modified Bessel function of the second kind, also known as the Macdonald function~\cite{debnath2016integral, abramowitz72}, of purely imaginary order $i\nu$, expressed in the form of an improper integral as~\cite{erdelyi54, prudnikov1992integrals}
\begin{equation}
    K_{i\nu}(\zeta) = \int_0^\infty 
    e^{-\zeta\ch t} \cos(\nu t) \, \mathrm{d}t \, .
    \label{eq:Kinu_def}
\end{equation}
It follows that $K_{i\nu}(\xi)$ is real for a positive argument $\xi$ when $\nu$ is real and is an even function of its index.

\begin{table*}[t]
    \centering
    {
\renewcommand{\arraystretch}{3.2}
\setlength{\tabcolsep}{15pt}
    \begin{tabular}{|c|c|}
    \hline
        $f$ & $\displaystyle  \Tinu \{f\} =  
    \int_{-\infty}^\infty \mathrm{d}z \, e^{ikz} \int_0^\infty f(r,z) \Kinukr \, r^{-1} \, \mathrm{d} r $ \\
    \hline
    \hline
    $\displaystyle \pder{f}{z}$ & $-ik \Tinu \{f\}$ \\
    \hline
    $\displaystyle \pder{f}{r}$ &
    $\displaystyle \frac{|k|}{2i\nu} 
    \left( (i\nu+1) \TinuP \{f\}
    +  (i\nu-1) \TinuN \{f\}
    \right) $ \\
    \hline
    $\displaystyle \frac{f}{r}$ & $\displaystyle \frac{|k|}{2i\nu} \left( \TinuP \{f\} - \TinuN \{f\} \right)$ \\
    \hline
    $\displaystyle \frac{z}{r}\, f$ &
    $\displaystyle -\frac{\sgn k}{2\nu} 
    \left(
    \left(
    i\nu + 1 + k\, \pder{}{k} \right) \TinuP \left\{ f\right\}
    + \left(
    i\nu - 1 - k\, \pder{}{k} \right) \TinuN \left\{ f\right\}
    \right)$ \\
    \hline 
    $r^2 \bm{\nabla}^2 f$ & $\displaystyle \left( \frac{\partial^2}{\partial\theta^2}-\nu^2 \right) \Tinu \{f\}$ \\
    \hline
    \hline
    \end{tabular}
    }
   \caption{Key properties of the FKL transform relevant for solving Stokes flow problems, including derivatives with respect to $r$ and $z$, operations such as division by $r$ and multiplication by $z/r$, and the transform of $r^2 \bm{\nabla} f$.
   Here, sgn denotes the sign function, defined as $\operatorname{sgn} k = k/|k|$.}
    \label{tab:FKL_properties}
\end{table*}

Throughout this paper, the hat symbol denotes the Fourier transform of the function of interest—here, the velocity field or pressure—while a tilde indicates the KL transform of the Fourier-transformed function; that is, the Fourier transform is performed first, followed by the KL transform. 
This order of integration is typically preferred, as the Fourier transform is often simpler to evaluate first.
Again, the polar angle remains unchanged under these transformations. Consequently, the partial differential equation in Eq.~\eqref{eq:Stokes}, which governs the hydrodynamic fields, is reduced to a system of ordinary differential equations in~$\theta$.

For convenience, we introduce the notation for the combined FKL transform as
\begin{equation}
    \label{eq:FKL_combined_notation}
    \widetilde{f} := 
    \Tinu \left\{ f \right\}
    = 
    \mathscr{K}_{i\nu} \left\{ \mathscr{F} \left\{ f \right\} \right\} .
\end{equation}
In addition, we define the abbreviation
\begin{equation}
    \Tinu^{\,\prime} \{f\}
    = \pder{}{\theta} \, \Tinu \{f\} \, .
\end{equation}

Next, we consider the inverse transforms, beginning with the inverse KL transform of Eq.~\eqref{eq:KL_forward}, which is given by
\begin{equation}
     \hat{f}(r,k) 
    = \frac{2}{\pi^2} 
    \int_0^\infty \widetilde{f}(\nu, k) \, \Kinukr \sh (\pi\nu) \, \nu \, \mathrm{d} \nu \, . \label{eq:inv_KL}
\end{equation}

Applying the inverse Fourier transform to Eq.~\eqref{eq:inv_KL} yields
\begin{equation}
   f(r, z) 
    = \frac{1}{2\pi}
    \int_{ -\infty }^{\infty} 
    \hat{f}(r,k) \, e^{-ikz} \, \mathrm{d} k \, .
     \label{eq:inv_F}
\end{equation}

It should be emphasized that, in practice, the inverse transformation is rarely performed by applying the inverse KL transform stated by Eq.~\eqref{eq:inv_KL} first. Instead, the inverse Fourier transform is typically evaluated initially, as many of the resulting integrals with respect to the axial wavenumber~$k$ can be conveniently computed using standard tabulated formulas. This procedure reduces the problem to a single infinite integral over the radial wavenumber~$\nu$, which is generally handled either numerically or, in certain cases, analytically via the method of residues.

As $\zeta \to \infty$, the modified Bessel function of the second kind defined by Eq.~\eqref{eq:Kinu_def} exhibits the following asymptotic behavior~\cite{lebedev1972special, yakubovich1996index}
\begin{equation}
    K_{i\nu}(\zeta) = \sqrt{\frac{\pi}{2\zeta}} \, 
    e^{-\zeta} \left( 1+\mathcal{O} \left( \frac{1}{\zeta} \right) \right) .
    \label{eq:BesselK_large_arg}
\end{equation}
For $|\nu| \to \infty$ with $\xi > 0$ fixed, we have~\cite{lebedev1972special, yakubovich1996index}
\begin{equation}
    K_{i\nu}(\xi) = \mathcal{O}
    \left( 
    \frac{e^{-\frac{\pi}{2} |\nu|}}{\sqrt{|\nu|}} 
    \right) .
    \label{eq:BesselK_large_index}
\end{equation}

The asymptotic behavior of $K_{i\nu}$ indicates that the function decays rapidly for both large arguments and large orders. The exponential decay as $\zeta \to \infty$ ensures that contributions from distant regions are negligible, while the exponential decay as $|\nu| \to \infty$ suppresses high-order modes. These properties justify truncating improper integrals in practical computations, improve numerical stability, and reflect the localization of effects in physical problems, such as flows near boundaries.

\subsection{Properties of the FKL Transform}
    
\subsubsection{Functional relationships}

Here, we summarize key properties of the FKL transform that are relevant for solving Stokes flow problems involving a point force or point torque singularity. These include the FKL transform of derivatives with respect to $r$ and $z$, operations such as division by $r$ and multiplication by $z/r$, as well as the FKL transform of the Laplace equation. Proofs of these properties are provided in the Appendix of Ref.~\onlinecite{daddi2025proc}, so they are not repeated here. The results are presented in Tab.~\ref{tab:FKL_properties} to facilitate easy reference for the reader.

When the transformed function~$f$ is an even function of $z$, as is the case here, its FKL transform is real. Consequently, for transforms that preserve $z$-parity, the FKL transform remains real, and the two terms defining the transform of $\pderline{f}{r}$ and $f/r$ are complex conjugates, allowing the transform to be obtained by taking twice the real part of either term. In contrast, for transforms that reverse $z$-parity, the transform of an even function is purely imaginary. In this case, the two terms of the transforms of $\pderline{f}{z}$ and $(z/r) f$ are negative complex conjugates, with identical imaginary parts and real parts equal in magnitude but opposite in sign. 
We chose to present the results as shown in Tab.~\ref{tab:FKL_properties}, as this allows for a more streamlined derivation when applying the properties of the FKL transform.

\subsubsection{FKL transform of $1/s$}

In the context of three-dimensional Stokes flow induced by a force or torque singularity, the FKL transform of $1/s$ is a fundamental quantity that helps construct the free-space contribution to the solution.
For a point force, the harmonic functions are expressed directly in terms of $1/s$, whereas for a point torque they are given in terms of partial derivatives of $1/s$.

In cylindrical coordinates, $s$ takes the form
\begin{equation}
    \label{eq:s_def}
    s := |\R-\R_0| = \sqrt{r^2+\rho^2+z^2-2\rho r \cos(\theta-\beta)} \, . 
\end{equation}

As outlined above, the determination of the FKL transform proceeds by first evaluating the Fourier transform, followed by the KL transform.
We determine the Fourier transform of $1/s$ as defined by Eq.~\eqref{eq:Fourier_forward}. Taking into account the parity with respect to $z$, we have
\begin{equation}
    \mathscr{F} \left\{ \invs \right\}
    = 2 \int_0^\infty \frac{\cos(kz) }{\sqrt{z^2 + s_0^2}} \, \mathrm{d}z \, ,
    \label{eq:inv_s_Fourier_tmp}
\end{equation}
where we have defined
\begin{equation}
    s_0 = s(z=0) = \sqrt{r^2+\rho^2-2\rho r\cos(\theta-\beta)} \, .
\end{equation}

We invoke the integral representation of the zeroth-order modified Bessel function of the second kind, see Watson~\cite[p.~183]{watson22}, or Gradshteyn and Ryzhik~\cite[p.~435]{gradshteyn2014table}, Eq.~3.754.2:
\begin{equation}
    \label{eq:watson}
    K_0(b\zeta) = \int_{0}^\infty 
    \frac{\cos(b t)}{\sqrt{t^2+\zeta^2}}\,\mathrm{d}t \, , 
\end{equation}
for $b>0$ and $\operatorname{Re} \zeta>0\, .$

The integral in Eq.~\eqref{eq:inv_s_Fourier_tmp} can be evaluated by substituting $b=|k|$ and $\zeta=s_0$ into Eq.~\eqref{eq:watson}, yielding
\begin{equation}
   \mathscr{F} \left\{ \invs \right\}
    = 2 K_0(|k|s_0) \, .
    \label{eq:inv_s_Fourier}
\end{equation}

We next employ the classic textbook \textit{Table of Integral Transforms} by Erdélyi \textit{et al.}~\cite[p.~175]{erdelyi54} to derive the KL transform of Eq.~\eqref{eq:inv_s_Fourier}.
Specifically,
\begin{equation}
    \hspace{-0.2cm}
    \int_0^\infty  K_0 \left( q(y) \right) K_{ix} (y) \, y^{-1} \, \mathrm{d}y 
    = \frac{\pi}{x \sh (\pi x)} \ch(ax) K_{i x}(b) \, ,
    \label{eq:erdelyi_int}
\end{equation}
where we have defined the abbreviation
\begin{equation}
    q(y) = \sqrt{y^2+b^2+2by\cos a} \, .
\end{equation}
For general complex numbers $a$ and $b$, the integral given in Eq.~\eqref{eq:erdelyi_int} is defined when $|\operatorname{Re} a| + |\operatorname{arg} b| < \pi$.

Finally, by applying the changes of variables $x=\nu$, $y=|k|r$, and $b=|k|\rho$, the FKL transform of $1/s$ reads
\begin{equation}
    \label{eq:inv_s_FKL}
    \Tinu \left\{ \invs \right\}
    = 
    \frac{2\pi}{\nu \sh(\pi\nu)} \ch \left( a \nu \right) K_{i\nu} (|k|\rho) \, ,
\end{equation}
where
\begin{equation}
    a = \pi-|\theta-\beta| \, .
    \label{eq:a_def}
\end{equation}
This result constitutes a central element of the FKL method when applied to three-dimensional Stokes flow near wedges.
Here, $a \in (0,\pi] \, .$

\begin{table}[]
    \centering
    {
    \renewcommand{\arraystretch}{1.5}
    \setlength{\tabcolsep}{10pt}
    \begin{tabular}{|c|c|c|c|c|}
       \hline
       Coefficients & $F_\parallel$ & $F_\perp$ & $T_\parallel$ & $T_\perp$ \\
       \hline
       \hline 
       $\Lambda_x, \, \Lambda_x^\dagger$ & ~ & \checkmark & \checkmark & ~ \\
       $\Lambda_y, \, \Lambda_y^\dagger$ & ~ & \checkmark & \checkmark & ~ \\
       $\Lambda_z, \, \Lambda_z^\dagger$ & \checkmark & ~ & ~ & \checkmark \\
       $\Lambda_w, \, \Lambda_w^\dagger$ & ~ & \checkmark & \checkmark & ~ \\ 
       \hline
       $\operatorname{H}_x, \, \operatorname{H}_x^\dagger$ & ~ & \checkmark & \checkmark & ~ \\
       $\operatorname{H}_y, \, \operatorname{H}_y^\dagger$ & ~ & \checkmark & \checkmark & ~ \\
       $\operatorname{H}_z, \, \operatorname{H}_z^\dagger$ & ~ & ~ & ~ & \checkmark \\
       \hline
       $\Delta_x, \, \Delta_x^\dagger$ & \checkmark & ~ & ~ & \checkmark \\
       $\Delta_y, \, \Delta_y^\dagger$ & \checkmark & ~ & ~ & \checkmark \\
       $\Delta_w, \, \Delta_w^\dagger$ & ~ & ~ & ~ & \checkmark \\
       \hline
       \hline
    \end{tabular}
    }
    \caption{Nonzero coefficients defining $A_j$ and $A_j^\dagger$ for various orientations of point forces and torques; c.f.\ Eq.~\eqref{eq:A_j_general_expr}.
    A transverse force and a parallel torque exhibit the same pattern of non-vanishing coefficients, as required.
    }
    \label{tab:coeffs}
\end{table}

\subsection{Complementary solution}

\subsubsection{Representation of the solution in FKL space}

The four harmonic functions $\phi_j$, $j \in \{x,y,z,w\}$, defined in Eq.~\eqref{eq:solution_decompo} and representing the complementary solution, satisfy the Laplace equation.
As shown in the last entry of Tab.~\ref{tab:FKL_properties}, the Laplace equation takes a remarkably simple form in FKL space, reducing to a linear second-order ordinary differential equation with constant coefficients, whose solution is given by
\begin{equation}
    \label{eq:phi_SOLUTION_FORM}
    \phitil_j(\nu,\theta,k) = A_j(\nu,k) \sh(\theta\nu) + A^\dagger_j(\nu,k) \ch(\theta\nu) \, , 
\end{equation}
where $A_j$ and $A_j^\dagger$ are unknown wavenumber-dependent functions to be determined from the boundary conditions.

\begin{table*}[t]
    \centering
    {
\renewcommand{\arraystretch}{3.5}
\setlength{\tabcolsep}{15pt}
    \begin{tabular}{|c|c|}
    \hline
    $f$ & $\displaystyle  \Tinu \{f\} =  
    \int_{-\infty}^\infty \mathrm{d}z \, e^{ikz} \int_0^\infty f(r,z) \Kinukr \, r^{-1} \, \mathrm{d} r $ \\
    \hline
    \hline
    $\displaystyle \frac{z}{r} \pder{}{a} \invs$ & $\displaystyle ik\rho \sin a \, \Tinu \left\{ \invs \right\}$ \\
    \hline
    $\displaystyle \frac{\rho}{r} \pder{}{a} \invs$ &
    $\displaystyle -\left( \cos a \, \pder{}{a} + \sin a \, \rho\, \pder{}{\rho} \right) \Tinu \left\{ \invs \right\}$ \\
    \hline
    $\displaystyle \rho\, \pder{}{r} \invs$ &
    $\displaystyle -\left( \left( \nu^2\sin a + \cos a \, \pder{}{a} \right)
    + \rho \, \pder{}{\rho} 
    \left( \sin a  - \cos a \, \pder{}{a} \right) \right) \Tinu \left\{ \invs \right\} $ \\
    \hline
    \end{tabular}
    }
    \caption{Expressions of the FKL transforms of various quantities relevant to the solution of a point torque, expressed in terms of the FKL transform of $1/s$ given by Eq.~\eqref{eq:inv_s_FKL} and its derivatives with respect to $a$ and $\rho$.
    We recall that $a=\pi-|\theta-\beta|$.
    }
    \label{tab:FKL-1-over-s}
\end{table*}

Typically, the expressions for $A_j$ take the general form
\begin{equation}
    A_j = \frac{2\pi}{\nu\sh(\pi\nu)} 
    \left( \Lambda_j + \operatorname{H}_j \rho \, \pder{}{\rho} + ik\rho \, \Delta_j \right) K_{i\nu} (|k|\rho) \, ,
    \label{eq:A_j_general_expr}
\end{equation}
for $j \in \{x,y,z,w\}$, where $\Lambda_j$, $\operatorname{H}_j$, and $\Delta_j$ are dimensionless quantities depending on $\nu$, $\alpha$, and $\beta$, and for transverse forces and torques also on their orientation. 
A similar expression to Eq.~\eqref{eq:A_j_general_expr} applies to $A^\dagger_j$, with $\Lambda_j$, $\operatorname{H}_j$, and $\Delta_j$ replaced by $\Lambda_j^\dagger$, $\operatorname{H}_j^\dagger$, and $\Delta_j^\dagger$, respectively.

The purpose of this general form 
is to significantly facilitate the inverse Fourier transform when obtaining the solution in real space.
Table~\ref{tab:coeffs} lists the non-vanishing coefficients defining $A_j$ and $A_j^\dagger$ for an axial point force~$F_\parallel$, a transverse point force~$F_\perp$, an axial point torque~$T_\parallel$, and a transverse point torque~$T_\perp$.
A transverse force and a parallel torque share the same set of non-vanishing coefficients, as expected.

For the transverse force, we define $\sigma$ as the angle between the force direction and the radial axis, i.e., $\sigma=\tan^{-1}(F_r/F_\theta)$, where $F_r$ and $F_\theta$ denote the radial and azimuthal components of the transverse force, respectively.
Accordingly, $\sigma=0$ corresponding to a purely radial force and $\sigma=\pi/2$ to a purely azimuthal force. The same definition applies to the angle for a transverse torque normal to the wedge edge.
Thanks to the linearity of the governing equations, any general motion can be described by a linear superposition of the solutions for axial and transverse singularities.

\subsubsection{Representation of the solution in real space}

By combining the inverse Fourier and inverse KL transforms, Eqs.~\eqref{eq:inv_KL} and \eqref{eq:inv_F}, the solution for the four harmonic functions can be expressed in real space as a double integral over the radial and axial wavenumbers. The integration over $k$ can be performed using tabulated integrals, leaving a single integral over $\nu$ that can be evaluated numerically.

Using the general form in Eq.~\eqref{eq:phi_SOLUTION_FORM} together with the representation in Eq.~\eqref{eq:A_j_general_expr}, the solution in real space can be expressed in the following integral form
\begin{equation}
    \phi_j (r, \theta, z) = \int_0^\infty 
   \Psi_j(\theta, \nu) \, \mathcal{K}_{i\nu}(r,z) \, \mathrm{d}\nu \, , 
    \label{eq:phi_parallel}
\end{equation}
where we have defined the kernel function
\begin{equation}
    \mathcal{K}_{i\nu}(r,z)
    = \frac{4}{\pi^2} \int_0^\infty  K_{i\nu}(k\rho) K_{i\nu}(kr) \cos(kz) \, \mathrm{d}k \, .
    \label{eq:Kinu_kernel}
\end{equation}
In addition,
\begin{equation}
    \Psi_j (\theta, \nu) =  \psi_j(\theta, \nu) +  \xi_j(\theta, \nu) \, \rho \, \pder{}{\rho}
    - \varphi_j(\theta, \nu) \, \rho \, \pder{}{z} \, , 
\end{equation}
where
\begin{subequations}
    \begin{align}
    \psi_j(\theta, \nu) &= \Lambda_j  \sh(\theta\nu) + \Lambda_j^\dagger  \ch(\theta\nu) \, , \\
    \xi_j(\theta, \nu) &= \mathrm{H}_j  \sh(\theta\nu) + \mathrm{H}_j^\dagger \ch(\theta\nu) \, , \\
     \varphi_j(\theta, \nu) &= \Delta_j  \sh(\theta\nu) + \Delta_j^\dagger \ch(\theta\nu) \, .    
\end{align}
\end{subequations}

The improper integral in Eq.~\eqref{eq:Kinu_kernel} converges, and its value is provided in classic textbooks as
\begin{equation}
    \mathcal{K}_{i\nu} (r,z) = 
    \frac{P_{i\nu-\frac{1}{2}} (c)}{\sqrt{\rho r}} \, \sch(\pi\nu) \, ,
    \label{eq:Kinu_result}
\end{equation}
see, for instance, the standard table of integrals by Gradshteyn and Ryzhik~\cite[p.~719]{gradshteyn2014table}, Eq.~6.672.3, or Prudnikov \textit{et al.}~\cite[p.~390]{prudnikov1992integrals}, Eq.~2.16.36.2. Here, $P_n$ denotes the Legendre function of the first kind of degree $n$, with positive argument
\begin{equation}
    c = \frac{1}{2\rho r} \left( \rho^2+r^2+z^2 \right) .
    \label{eq:c}
\end{equation}

In the following, we present the methodological steps leading to the system of equations whose solution determines the coefficients $A_j$ and $A_j^\dagger$. 
Some lengthy expressions associated with the solution for a point torque are omitted here, as they can be readily obtained from the details provided here, or found in Ref.~\onlinecite{daddi2026hydrodynamic}.

In the limit $r\to 0$, relevant for describing the flow behavior near the apex of the wedge, the kernel function given by Eq.~\eqref{eq:Kinu_result} assumes the asymptotic form, cf.\ Appendix
\begin{equation}
    \hspace{-0.25cm}
    \mathcal{K}_{i\nu} (r,z) \simeq 
    \frac{2\csh(\pi\nu)}{\sqrt{\pi}}
    \operatorname{Im}
    \left\{
    \frac{ \Gamma \left( \frac{1}{2}-i\nu \right) }{ \Gamma \left( 1-i\nu \right) }
    \frac{ \left( \rho^2+z^2 \right)^{i\nu-\frac{1}{2}} }{ \left( \rho r\right)^{i\nu} }
    \right\} 
    \label{eq:Kinu_result_asym}
\end{equation}
with $\Gamma$ denoting the Euler gamma function.

The asymptotic expression of $\phi_j$ as $r\to 0$ can be obtained by performing the integration over $\nu$ using the method of residues. The integration can be performed in the upper half of the complex plane using a contour composed of the real axis and a large counterclockwise semicircular arc of radius~$R$, with~$R$ taken to infinity. In some cases, a small clockwise contour around the origin must be included when zero is a pole. Since there are infinitely many poles, the leading-order behavior is captured by considering only the first few.
Not all poles can be determined analytically, which prevents a complete asymptotic solution for an arbitrary wedge opening angle.

\section{Solution for a point force or a point torque}
\label{sec:solution}

\subsection{Parallel point force}

In an unbounded medium, the solution for a point force applied along the $z$-direction, parallel to the wedge edge, is given by
\begin{equation}
    \phi_r^\infty = \phi_\theta^\infty = \phi_w^\infty = 0 \, , \quad
    \phi_z^\infty = -\invs \, .  
\end{equation}

In the FKL space, $\phitil_z^{\,\infty}$ can directly be obtained from Eq.~\eqref{eq:inv_s_FKL}.
By enforcing zero radial and axial velocities on the wedge surfaces, it follows from Eq.~\eqref{eq:NS_BCs} that the complementary solution must satisfy
\begin{equation}
    \phitil_r = 0 \, , \qquad \phitil_w = 0 \, , \qquad \phitil_z  + \phitil_z^{\,\infty} = 0 \, , 
\end{equation}
evaluated at $\theta = \pm\alpha$. 
Since $\phitil_w$ vanishes on the wedge surfaces, it follows from Eq.~\eqref{eq:phi_SOLUTION_FORM} that $\phitil_w = 0$ everywhere. 
However, this does not hold for $\phitil_r$, since $\phitil_r=0$ at $\theta=\pm\alpha$ does not imply that it vanishes everywhere, as it is not a harmonic function; see Eq.~\eqref{eq:Phi_r_Phi_The}, from which the expressions for $\phitil_r$ and $\phitil_\theta$ can be obtained.

By imposing a zero azimuthal velocity boundary condition on the wedge surfaces, c.f.\ Eq.~\eqref{eq:NS_BCs}, we require that
\begin{equation}
   \Tinu^{\,\prime}
    \left\{
    \phi_r +
    \frac{z}{r} \left( \phi_z + \phi_z^{\infty} \right) \right\} 
    - 2 \,\phitil_\theta 
    = 0 \, ,  
    \label{eq:force-axial-1-eqn}
\end{equation}
evaluated at $\theta = \pm\alpha\, .$ 

As shown in Tab.~\ref{tab:coeffs}, for an axial point force, $A_z$ is expressed in terms of $\Lambda_z$, whereas $A_x$ and $A_y$ are expressed in terms of $\Delta_x$ and $\Delta_y$, respectively. 
The same applies to the corresponding quantities with a dagger.

The solution for $\Lambda_z$ and $\Lambda_z^\dagger$ follows directly from the equation $\phitil_z  + \phitil_z^{\,\infty} = 0$ imposed at $\theta=\pm\alpha$, yielding two equations whose solutions are given by
\begin{subequations}
    \begin{align}
    \Lambda_z &= \sh(\beta\nu) \sh ( \overline{\alpha}\nu) \csh (\alpha\nu)\, , \\
    \Lambda_z^\dagger &= \ch(\beta\nu) \ch ( \overline{\alpha}\nu) \sch (\alpha\nu) \, , 
\end{align}
\end{subequations}
where the supplementary angle of $\alpha$ is defined as $\overline{\alpha} = \pi - \alpha$.

The expressions for $\Delta_j$ and $\Delta_j^\dagger$, $j\in\{x,y\}$, are obtained by solving the system of four equations resulting from $\phitil_r=0$ together with Eq.~\eqref{eq:force-axial-1-eqn}, evaluated at $\theta=\pm\alpha$.
To evaluate the FKL transform of $(z/r) f$, we use the property of FKL transforms listed in Tab.~\ref{tab:FKL_properties}.
Considering the free-space contribution obtained in FKL space from Eq.~\eqref{eq:inv_s_FKL}, we use the first entry of Tab.~\ref{tab:FKL-1-over-s} to obtain
\begin{equation}
    \Tinu^{\,\prime}
    \left\{ \frac{z}{r}\, \phi_z^{\infty} \right\}
    = ik\rho \, \phitil_z^{\,\infty} \, \sin(\beta-\theta) \, .
\end{equation}
The expressions for the coefficients $\Delta_j$ and $\Delta_j^\dagger$, for $j\in \{x,y\}$ are obtained as
\begin{equation}
    \Delta_x = h_1 \sin\alpha \ch (\alpha\nu)  \, , \quad
    \Delta_y =  h_2 \cos\alpha \ch (\alpha\nu) \, ,
\end{equation}
together with
\begin{equation}
    \Delta_x^\dagger = -h_2 \sin\alpha\sh(\alpha\nu)  \, , \quad
    \Delta_y^\dagger = -h_1 \cos\alpha\sh(\alpha\nu)  \, ,
\end{equation}
where \( h_1 \) and \( h_2 \) are given by
\begin{subequations}
    \begin{align}
    h_1 &= 4\Gamma_+ \sh(\pi\nu) \big( \cos\alpha\sin\beta\sh(\alpha\nu)\ch(\beta\nu) \notag \\
    &\quad-\sin\alpha\cos\beta\ch(\alpha\nu)\sh(\beta\nu) \big)  \, ,\\[3pt]
    h_2 &= 4\Gamma_- \sh(\pi\nu) \big( \sin\alpha\cos\beta\sh(\alpha\nu)\ch(\beta\nu) \notag \\
    &\quad-\cos\alpha\sin\beta\ch(\alpha\nu)\sh(\beta\nu) \big) \, ,
\end{align}
\end{subequations}
where
\begin{equation}
    \hspace{-0.1cm}
    \Gamma_\pm^{-1} = \left( \sh(2\alpha\nu)\pm \nu \sin(2\alpha) \right)
    \left( \ch (2\alpha\nu) \mp \cos(2\alpha) \right) .
\end{equation}

\subsection{Transverse point force}

In an unbounded fluid medium, the velocity field induced by a point-force singularity applied in the direction normal to the edge of the wedge is derived from
\begin{subequations}
   \begin{align}
        \phi_x^\infty &= -\invs \cos(\beta + \sigma) \, , \\
        \phi_y^\infty &= -\invs \sin(\beta + \sigma) \, , \\
        \phi_w^\infty &= \invs \cos\sigma \, , 
   \end{align}
\end{subequations}
and $\phi_z^\infty = 0\, .$
The FKL transforms of the harmonic functions $\phi_j^{\infty}$, for $j\in\{x,y,w\}$, can be readily obtained using Eq.~\eqref{eq:inv_s_FKL}.

By imposing vanishing radial and axial velocities on the wedge surfaces, Eq.~\eqref{eq:NS_BCs} implies that
\begin{equation}
    \phitil_r + \phitil_r^{\,\infty} = 0 \, , \quad
    \phitil_z = 0 \, , \quad
    \phitil_w + \phitil_w^{\,\infty} = 0  \, ,
\end{equation}
evaluated at $\theta = \pm\alpha \, .$
Because $\phitil_z$ vanishes on the wedge surfaces, Eq.~\eqref{eq:phi_SOLUTION_FORM} implies that $\phitil_z$ must be zero everywhere.

By requiring vanishing azimuthal velocity on the wedge surface, it follows that
\begin{equation}
    \Tinu^{\,\prime} \left\{
    \phi_r + \phi_r^{\,\infty} +
    \frac{\rho}{r} \left( \phi_w + \phi_w^\infty \right) 
    \right\}
    -2 \left( \phitil_\theta + \phitil_{\theta}^{\,\infty} \right) = 0 \, , 
    \label{eq:force-transverse-1-eqn}
\end{equation}
imposed at $\theta = \pm\alpha \, .$

As shown in Tab.~\ref{tab:coeffs}, for a transverse point force or an axial point torque, $A_w$ is expressed in terms of $\Lambda_w$, while $A_j$, $j\in\{x,y\}$, are expressed in terms of $\Lambda_j$ and $\operatorname{H}_j$.
The same holds for the corresponding daggered quantities.

The solution for $\Lambda_w$ and $\Lambda_w^\dagger$ follows directly from imposing $\phitil_w + \phitil_w^{\,\infty} = 0$ at $\theta=\pm\alpha$, resulting in two equations whose solutions are given by
\begin{subequations}
    \begin{align}
    \Lambda_w &= -\sh(\beta\nu) \sh ( \overline{\alpha}\nu) \csh (\alpha\nu) \cos\sigma \, , \\
    \Lambda_w^\dagger &= -\ch(\beta\nu) \ch ( \overline{\alpha}\nu) \sch (\alpha\nu) \cos\sigma  \, . 
\end{align}
\end{subequations}

The expressions for $\Lambda_j$, $\operatorname{H}_j$, $\Lambda_j^\dagger$, and $\operatorname{H}_j^\dagger$, $j\in\{x,y\}$, are obtained by solving the system formed by $\phitil_r + \phitil_r^{\,\infty} = 0$ and Eq.~\eqref{eq:force-transverse-1-eqn} evaluated at $\theta=\pm\alpha$.
For the evaluation of the FKL transform of $\phi_w/r$, we make use of Tab.~\ref{tab:FKL_properties}.
With the free-space contribution obtained in FKL space from Eq.~\eqref{eq:inv_s_FKL}, we use directly the second entry of Tab.~\ref{tab:FKL-1-over-s} to obtain
\begin{equation}
\hspace{-0.1cm}
    \Tinu^{\,\prime} \left\{ \frac{\rho}{r}\, \phi_w^\infty \right\}
    = \operatorname{sgn} (\theta-\beta) 
    \left( \cos a \, \pder{}{a} +  \rho \sin a\, \pder{}{\rho} \right) \phitil_w^{\,\infty} , 
\end{equation}
where $a$ is defined by Eq.~\eqref{eq:a_def}.
Defining the abbreviations $\zeta_\pm = \alpha \pm \beta$, 
$\xi_\pm = \alpha \pm \beta \pm \sigma$, 
$\beta_\pm = \pi \pm \beta$, and
$\tau = \beta+\sigma$, the final results can be represented in the form
\begin{subequations}
    \begin{align}
    \Lambda_x &= \Pi_+ \left( b \cos\tau + \nu \left( {a}_- + {c}_- \ch(\alpha\nu) \right) \sin\alpha \right) , \\[3pt]
    \Lambda_y &= \Pi_- \left( b \sin\tau + \nu \left( {a}_+ - {c}_+ \ch(\alpha\nu) \right) \cos\alpha \right) , \\[3pt]
    \Lambda_x^\dagger &= 
    \Pi_- \left( b^\dagger \cos\tau - \nu \left( {a}_+^\dagger - {c}_+ \sh(\alpha\nu) \right) \sin\alpha \right) , \\
    \Lambda_y^\dagger &= 
    \Pi_+ \left( b^\dagger \sin\tau - \nu \left( {a}_-^\dagger + {c}_- \sh(\alpha\nu) \right) \cos\alpha \right) .
\end{align}
\end{subequations}
The coefficients associated with $\rho \, \partial/\partial\rho$ in Eq.~\eqref{eq:A_j_general_expr} can be cast in the form
\begin{subequations}
    \begin{align}
\hspace{-0.2cm}
     \operatorname{H}_x &= - \Pi_+ d_- \sin\alpha \ch(\alpha\nu) \, , \,
    \operatorname{H}_y = \Pi_- d_+ \cos\alpha \ch(\alpha\nu) \, , \\
\hspace{-0.2cm}
    \operatorname{H}_x^\dagger &= -\Pi_- d_+ \sin\alpha \sh(\alpha\nu) \, , \,
    \operatorname{H}_y^\dagger = \Pi_+ d_- \cos\alpha \sh(\alpha\nu) \, , 
\end{align}
\end{subequations}
where
\begin{equation}
    \Pi_\pm^{-1} = \sh(2\alpha\nu) \pm \nu \sin(2\alpha) \, .
\end{equation}
We have introduced the following expressions to present the results more concisely
\begin{subequations}
    \begin{align}
    {a}_\pm &=  \cos\xi_+ \sh\left( \beta_- \nu\right) \pm \cos\xi_- \sh\left( \beta_+\nu\right) , \\
    {a}_\pm^\dagger &=  \cos\xi_+ \ch\left( \beta_- \nu\right) \pm \cos\xi_- \ch\left( \beta_+\nu\right) , 
\end{align}
\end{subequations}
together with
\begin{subequations}
    \begin{align}
        {c}_\pm &= Z_\pm \left( \cos\zeta_-\ch \left( \zeta_+\nu \right) \pm 
    \cos\zeta_+ \ch \left( \zeta_-\nu\right) \right) , \\
    d_\pm &= Z_\pm \left( \sin\zeta_- \, \sh \left( \zeta_+\nu \right) \pm 
    \sin\zeta_+ \,\sh \left( \zeta_-\nu\right) \right) ,
    \end{align}
\end{subequations}
where 
\begin{equation}
    Z_\pm = \frac{2\cos\sigma \sh(\pi\nu)}{ \ch\left( 2\alpha\nu\right) \pm \cos(2\alpha)} \, .
\end{equation}
Moreover, 
\begin{subequations}
    \begin{align}
    b &= 2 \ch(\alpha\nu) \sh (\overline{\alpha}\nu) \sh(\beta\nu) \, , \\
    b^\dagger &= 2 \sh(\alpha\nu) \ch (\overline{\alpha}\nu) \ch(\beta\nu) \, ,
    \end{align}
\end{subequations}
where again $\overline{\alpha}=\pi-\alpha\, .$

\subsection{Parallel point torque}

For the determination of the solution for a point torque, the following identities prove to be useful
\begin{subequations} 
    \label{eq:torque-identities}
    \begin{align}
    \pder{}{r} \invs &=
    \frac{1}{s^3} \left( \rho\cos(\theta-\beta)-r \right) , 
\\
\pder{}{z} \invs &= -\frac{z}{s^3} \, , 
\\
\pder{}{\rho} \invs &=
    \frac{1}{s^3} \left( r\cos(\theta-\beta)-\rho \right) , \\
\pder{}{a} \invs &=
    \frac{\rho r}{s^3} \,\sin a \, .
\end{align}
\end{subequations}

The idea is to leverage the known FKL transform of $1/s$ by expressing the free-space solution in terms of partial derivatives. While derivatives with respect to $\rho$ or $\theta$ do not affect the FKL transform, the properties of the FKL transform must be applied when performing operations involving $r$ or $z$, or taking partial derivatives with respect to $r$ and $z$.

In an infinite medium, the solution for a point torque oriented along the $z$-direction is given by
\begin{subequations}
    \begin{align}
    \phi_r^\infty &= \frac{\rho^2}{s^3} \sin(\theta-\beta) \, , \\
    \phi_\theta^\infty &= \frac{\rho}{s^3} \left( \rho\cos(\theta-\beta)-r \right) \, , \\
    \phi_w^\infty &= -\frac{\rho r}{s^3}  \sin(\theta-\beta) \, .
\end{align}
\end{subequations}
and $\phi_z^\infty = 0$.
In addition, 
\begin{equation}
    r\phi_r^\infty + \rho \phi_w^\infty = 0 \, .
\end{equation}

Using the identities given in Eq.~\eqref{eq:torque-identities}, the free-space solution can be expressed in FKL space as
\begin{subequations}
   \begin{align}
        \phitil_r^{\,\infty} &= \sgn (\theta-\beta) 
         \, \Tinu \left\{ \frac{\rho}{r} \pder{}{a} \invs  \right\} , \\
         \phitil_\theta^{\,\infty} &= \Tinu \left\{ \rho\, \pder{}{r} \invs \right\}  , \\
        \phitil_w^{\,\infty} &= - \sgn (\theta-\beta) \, 
        \pder{}{a} \, \Tinu \left\{  \invs \right\} .
   \end{align}
\end{subequations}
The expressions for the free-space contribution can be obtained in FKL space using Tab.~\ref{tab:FKL-1-over-s}, where they are expressed in terms of derivatives with respect to $a$ and $\rho$ of the FKL transform of $1/s$ given in Eq.~\eqref{eq:inv_s_FKL}.

The condition of vanishing radial and axial velocities on the wedge surfaces yields
\begin{equation}
    \phitil_r + \phitil_r^\infty = 0 \, , \quad
    \phitil_z = 0 \, , \quad
    \phitil_w  + \phitil_w^\infty = 0 \, ,
\end{equation}
evaluated at $\theta=\pm\alpha \, .$
It follows from the no-slip condition on both wedge surfaces that $\phitil_z$ vanishes everywhere.
The condition of zero velocity in the azimuthal direction leads to
\begin{equation}
    \Tinu^{\,\prime} \left\{ 
     \phi_r + \frac{\rho}{r} \, \phi_w \right\} 
    -2 \left( \phitil_\theta + \phitil_\theta^\infty \right) = 0 \, , 
    \label{eq:torque-axial-1-eqn}
\end{equation}
imposed at $\theta = \pm\alpha \, .$

As noted above, for an axial point torque, $A_w$ is expressed in terms of $\Lambda_w$, while $A_j$, $j\in\{x,y\}$, are expressed in terms of $\Lambda_j$ and $\operatorname{H}_j$ (cf. Tab.~\ref{tab:coeffs}). The same applies to the corresponding quantities with a dagger.

The solution for the coefficients $\Lambda_w$ and $\Lambda_w^\dagger$ is obtained by imposing $\phitil_w + \phitil_w^{\,\infty} = 0$ at $\theta=\pm\alpha$, yielding two equations with solutions given by
\begin{subequations}
    \begin{align}
    \Lambda_w &= \nu\ch(\beta\nu) \sh \left( \overline{\alpha}\nu\right) \csh(\alpha\nu) \, , \\
    \Lambda_w^\dagger &= \nu\sh(\beta\nu) \ch \left( \overline{\alpha}\nu\right) \sch(\alpha\nu) \, .
\end{align}
\end{subequations}

The expressions for $\Lambda_j$, $\operatorname{H}_j$, $\Lambda_j^\dagger$, and $\operatorname{H}_j^\dagger$, $j\in\{x,y\}$, are obtained by solving the system composed of $\phitil_r + \phitil_r^{\,\infty} = 0$ and Eq.~\eqref{eq:torque-axial-1-eqn} evaluated at $\theta=\pm\alpha$.
To evaluate the FKL transform of $\phi_w/r$, we use the properties listed in Tab.~\ref{tab:FKL_properties}.
These coefficients are provided in Eq.~(32) of Ref.~\onlinecite{daddi2026hydrodynamic}.

\subsection{Transverse point torque}

The solution in an unbounded fluid medium for a transverse rotlet is given by
\begin{subequations}
    \begin{align}
    \phi_r^\infty &= \frac{z}{s^3}\, \sin( \theta-\beta-\sigma) \, , \\
    \phi_\theta^\infty &= \frac{z}{s^3}\, \cos(\theta-\beta-\sigma) \, , \\
    \phi_z^\infty &= -\frac{1}{s^3}
    \left( r\sin(\theta-\beta-\sigma) + \rho\sin\theta  \right) , \\
    \phi_w^\infty &= \frac{z}{s^3}\, \sin\sigma \, .
\end{align}
\end{subequations}
By construction
\begin{equation}
    r\phi_r^\infty + z\phi_z^\infty + \rho \phi_w^\infty = 0\, .
\end{equation}
Using the identities given in Eq.~\eqref{eq:torque-identities}, the free-space solution may be expressed in FKL space in the form
\begin{subequations}
    \begin{align}
    \phitil_r^\infty &= - \rho \sin(\theta-\beta-\sigma) \,  
    \Tinu \left\{ \pder{}{z} \invs \right\} , 
    \\[1pt]
    \phitil_\theta^\infty &= - \rho \cos(\theta-\beta-\sigma) \, 
    \Tinu \left\{ \pder{}{z} \invs \right\} , 
    \\[1pt]
    \phitil_z^\infty &=  \left( \sin\sigma \, \rho \, \pder{}{\rho}  -\cos\sigma \sgn(\theta-\beta) \,
    \pder{}{a} \right) \Tinu\left\{ \invs \right\} , \\[1pt]
    \phitil_w^\infty &= - \rho \sin\sigma \, \Tinu \left\{ \pder{}{z} \invs \right\}.
\end{align}
\end{subequations}

The conditions of vanishing radial and axial velocities on the wedge surfaces imply
\begin{equation}
    \phitil_r + \phitil_r^{\,\infty}  = 0 \, , \quad
    \phitil_z + \phitil_z^{\,\infty}  = 0 \, , \quad
    \phitil_w + \phitil_w^{\,\infty}  = 0 \, ,
\end{equation}
imposed at $\theta=\pm\alpha$.
Meanwhile, the condition of vanishing azimuthal velocity yields
\begin{equation}
    \mathscr{T}_{i\nu}^{\,\prime}
    \left\{ \phi_r + \frac{z}{r}\, \phi_z + \frac{\rho}{r}\, \phi_w \right\} - 2\left( \phitil_\theta + \phitil_\theta^\infty \right) = 0 \, , 
    \label{eq:torque-transverse-1-eqn}
\end{equation}
required at $\theta = \pm\alpha$.

As shown in Tab.~\ref{tab:coeffs}, $A_z$ is expressed in terms of $\Lambda_z$ and $\operatorname{H}_z$, whereas $A_j$, $j\in \{x,y,w\}$, are expressed in terms of $\Delta_j$ The same holds for the corresponding daggered quantities.

The solution for $\Delta_w$ and $\Delta_w^\dagger$ follows readily by solving $\phitil_w + \phitil_w^{\,\infty}=0$ evaluated at $\theta=\pm\alpha$, yielding
\begin{subequations}
    \begin{align}
    \Delta_w &= - \sh(\beta\nu) \sh\left( \overline{\alpha}\nu\right) \csh(\alpha\nu) \sin\sigma \, , \\
    \Delta_w^\dagger &= - \ch(\beta\nu) \ch\left( \overline{\alpha}\nu\right)\sch(\alpha\nu) \sin\sigma \, .
\end{align}
\end{subequations}
The condition $\phitil_z+\phitil_z^{\,\infty}=0$ enforced at $\theta=\pm\alpha$ yields
\begin{subequations}
    \begin{align}
    \Lambda_z &= \nu  \ch(\beta\nu) \sh \left( \overline{\alpha}\nu \right) \csh(\alpha\nu) \cos\sigma \, , \\
    \Lambda_z^\dagger &= \nu  \sh(\beta\nu) \ch \left( \overline{\alpha}\nu \right) \sch(\alpha\nu) \cos\sigma \, .
\end{align}
\end{subequations}
Moreover,
\begin{equation}
    \mathrm{H}_z = \Delta_w \, , \quad  \mathrm{H}_z^\dagger = \Delta_w^\dagger \, .
\end{equation}

The expressions for $\Delta_j$ and $\Delta_j^\dagger$, $j\in\{x,y\}$, can be obtained by solving the system of equations formed by $\phitil_r+\phitil_r^{\,\infty}=0$ together with Eq.~\eqref{eq:torque-transverse-1-eqn}, both evaluated at $\theta=\pm\alpha$.
We use the properties listed in Tab.~\ref{tab:FKL_properties} to evaluate the FKL transforms of $(z/r)\, \phi_z$ and $\phi_w/r$. 
These coefficients are provided in Eq.~(42) of Ref.~\onlinecite{daddi2026hydrodynamic}.

\section{Discussion and concluding remarks}
\label{sec:discussion}

\begin{figure}
    \centering
    \includegraphics[width=0.9\linewidth]{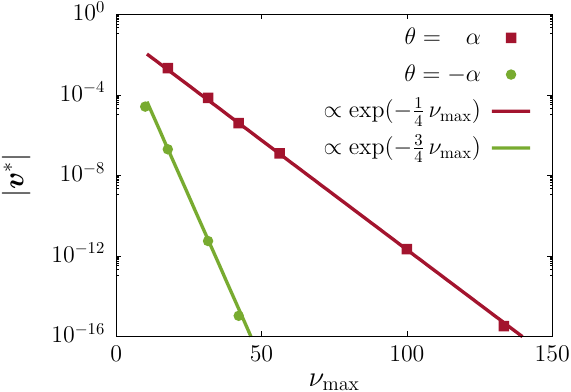}
    \caption{
    Scaled magnitude of the fluid velocity on the upper and lower walls induced by an axial point force, obtained via numerical integration with the upper limit truncated at $\nu_{\text{max}}$. The lines represent exponential fits. The velocity is scaled by $F/(8\pi\eta\rho)$. Here, $\alpha=\pi/6$, $\beta=\alpha/2$, and $r/\rho=z/\rho=1/2$.
    }
    \label{fig:NS}
\end{figure}

We presented an overview of the application of the FKL transform to address problems in fluid mechanics related to the fundamental solutions of the Stokes equations in wedge-shaped geometries. We detailed the analytical approach based on the Papkovich–Neuber representation of the solution for incompressible flow, involving four harmonic functions. The approach combines the use of a Fourier transform along the axial direction and a KL transform along the radial direction. Upon applying the FKL transform, the problem reduces to ordinary differential equations in the polar angle. The solution is expressed as the sum of the free-space contribution and a complementary solution required to satisfy the no-slip boundary conditions of vanishing velocity at the wedge surfaces. We showed that, depending on the orientation of the point force or point torque, the solution assumes a particular structure. The solution in real space is obtained by inverse transform, where the integration with respect to the axial wavenumber is carried out analytically, leaving a single improper integral in the radial wavenumber $\nu$.

The present review focuses primarily on the mathematical formulation of the problems under investigation, beginning with the relevant properties and functional relationships of the FKL transform. The fundamental building block for solving Stokes flow problems in wedge geometries due to singularities is the FKL transform of $1/s$, whose partial derivatives can be used to construct higher-order singularities such as a rotlet. This framework can also be applied to problems related to the dynamics of active microswimmers through the evaluation of the force dipole, source dipole, force quadrupole, and related singularities~\cite{spagnolie12, elgeti15, mathijssen2016hydrodynamics, ibrahim2016walls, daddi2019frequency, kurzthaler2021microswimmers}. The results presented here correspond to the case of no-slip boundary conditions imposed on both walls. The same method can be extended to examine free-slip or stress-free boundary conditions, as well as mixed combinations of different types of boundary conditions.

Since the kernel is a modified Bessel function of the second kind, which exhibits asymptotically exponential decay for large indices; c.f.\ Eq.~\eqref{eq:BesselK_large_index}, the numerical integration converges rapidly for large radial wavenumbers. Accordingly, the numerical integration is typically performed with the upper integration limit truncated at a specified maximum radial wavenumber, ensuring that the no-slip boundary conditions are satisfied while contributions from larger wavenumbers remain negligible. 
Figure~\ref{fig:NS} illustrates this exponential decay for a parallel point force using representative values for the wedge geometry. The boundary condition on the lower wall at $\theta=-\alpha$ is satisfied with a smaller truncation of the upper limit, owing to the faster exponential decay. Similar overall dynamics are observed for a point force aligned along the transverse direction or for a point torque. Typically, setting $\nu_\text{max}=100$ ensures an error below $10^{-10}$, making it suitable for evaluating the improper integrals.
Increasing the upper integration limit further does not change the results, making this choice an efficient balance between accuracy and computational cost. In the planar boundary limit, corresponding to a wedge opening angle of $\pi$, the integrals admit a closed analytical form, thereby recovering the previously known solutions for a point force~\cite{blake1971note} and a point torque~\cite{blake1974fundamental}.

The FKL transform is primarily a technical method for addressing flows near corners, providing an efficient framework for solving such problems. Unlike the method of successive reflections, which is a classical singularity-based approach relying on repeated image systems and often resulting in cumbersome series, the FKL approach expresses the solution in terms of rapidly converging integrals. This not only simplifies the computational implementation but also allows the effects of wedge boundaries to be incorporated directly into the kernel, offering a clear and compact representation of the flow induced by point singularities.
However, the method is subject to certain limitations. The approach is specifically tailored to wedge-shaped domains and relies on the separability of the geometry in cylindrical coordinates; extending it to more complex or irregular geometries would require substantial modifications. Additionally, the method is formulated for isolated point singularities; extending it to fully distributed forces, such as those around particles, would require additional treatments.
The method presented here can, in principle, be extended to related geometries with corner-like boundaries. For example, it could be adapted to study flows in finite wedges with bounded radial or axial extent. This can be achieved using dual integral equation formulations, which the author has previously applied to study the effects of finite-sized boundaries on various Stokes flows~\cite{daddi2020dynamics, daddi2020axisymmetric, daddi2022diffusiophoretic, daddi2021steady, daddi2023axisymmetric}. The main requirement is that the geometry allows a separation of variables compatible with the transform approach, so that the flow can be expressed in terms of harmonic functions. While the transforms and boundary conditions may become more involved in these cases, the framework remains applicable and provides a systematic way to construct solutions for singularity-driven flows in more complex confined geometries.

The solution outlined in the present work has the only approximation stemming from the numerical evaluation of the integral over $\nu$, which can be made arbitrarily accurate by choosing a sufficiently large upper limit and appropriate quadrature. To our knowledge, no other analytical or semi-analytical method exists for wedge-shaped geometries, as the combination of a point singularity and wedge boundaries makes closed-form solutions extremely challenging. In contrast, numerical approaches such as boundary element or finite-volume methods require discretization of the fluid domain or boundaries, involving two- or three-dimensional integrations, which are computationally more demanding. By reducing the problem to one-dimensional integrals over the radial wavenumber, our method achieves a significant gain in efficiency while maintaining high accuracy, which is particularly advantageous when evaluating multiple configurations or performing parameter studies in wedge geometries.

Considerable efforts have been devoted in the literature to solving the Stokes equations for fluid flows and the Navier–Cauchy equations for linear elasticity using boundary element methods. These methods rely on singular solutions to construct numerical solutions by discretizing the boundaries of the domain. In wedge-like geometries, this generally requires meshing both the surfaces of the walls and the objects within the domain, which can be computationally intensive.
By directly incorporating the Green’s functions outlined in this work, which satisfy the no-slip boundary conditions, into the boundary element formulation, the numerical implementation is greatly simplified. In this approach, the wedge walls do not need to be discretized, leading to a substantial reduction in computational cost. The solution detailed here can therefore be used to simulate particle dynamics in wedge-like geometries, as well as the behavior of inclusions in elastic media, providing a versatile tool for modeling in these complex geometries.

\begin{acknowledgments}
    I am grateful to Andreas M. Menzel for insightful discussions and collaboration on earlier studies that helped shape the perspectives presented in this review article.
\end{acknowledgments}

\section*{Author contributions}
The author has accepted responsibility for the entire content of this manuscript and approved its submission.

\section*{Conflict of interest}
The author declares no conflict of interest.

\section*{Data Availability Statement}
The article reviews theoretical approaches and does not present or report any data.


\appendix

\section*{Asymptotic expression of the kernel function in the limit $r\to 0$}

To determine the asymptotic expression of the kernel function given by Eq.~\eqref{eq:Kinu_result_asym}, we make use of the integral representation of the Legendre function~\cite{erdelyi54}
\begin{equation}
    P_{i \nu - \frac{1}{2}} \left( c \right) = \frac{\sqrt{2}}{\pi} \, \cth(\pi\nu) 
\int_{\operatorname{ach}c}^\infty \frac{\sin(\nu s)}{\sqrt{\cosh s - c}} \, \mathrm{d}s \, ,
\label{eq:Legendre_appendix}
\end{equation}
with $\operatorname{ach} = \operatorname{ch}^{-1}$ denoting the inverse hyperbolic cosine function.
We recall that $c$ is defined in terms of $r$, $z$, and $\rho$ by Eq.~\eqref{eq:c}.

Using Eq.~\eqref{eq:Legendre_appendix}, the kernel function given by Eq.~\eqref{eq:Kinu_result} ca be expressed as
\begin{equation}
    \mathcal{K}_{i\nu} (r,z) = 
    \frac{\sqrt{2}}{\pi} \frac{\csh(\pi\nu)}{\sqrt{\rho r}} 
\int_{\operatorname{ach}c}^\infty \frac{\sin(\nu s)}{\sqrt{\cosh s - c}} \, \mathrm{d}s \, .
\end{equation}

In the limit $r\to 0$, we have $c \simeq (\rho^2+z^2)/(2\rho r)$, so that $r\to 0$ corresponds to $c\to\infty$.

We next apply the change of variables $\ch s = c u$, so that
\begin{equation}
    \mathrm{d}s = \frac{c\, \mathrm{d} u}{\sqrt{c^2u^2-1}}
    \simeq \frac{\mathrm{d} u}{u} \, .
\end{equation}
It follows that
\begin{equation}
    \mathcal{K}_{i\nu} (r,z) \simeq \frac{2}{\pi} \frac{\csh(\pi\nu)}{ \sqrt{\rho^2+z^2} }
    \int_1^\infty
    \frac{ \sin \left( \nu \operatorname{ach}(cu) \right) }{u\sqrt{u-1}}\, 
    \mathrm{d}u\, .
\end{equation}
Using the asymptotic expansion $\operatorname{ach}(c u) \simeq \ln(2c) + \ln u$ and Euler's notation for the exponential function, it follows that
\begin{equation}
    \mathcal{K}_{i\nu} (r,z) \simeq 
    \frac{2}{\pi} \frac{\csh(\pi\nu)}{ \sqrt{\rho^2+z^2} }
    \operatorname{Im} \left\{
    (2c)^{i\nu} 
    \int_1^\infty 
    \frac{u^{i\nu}\, \mathrm{d}u}{u\sqrt{u-1}}
    \right\} .
\end{equation}
The integral can be represented using the beta function and, equivalently, expressed in terms of the gamma function as
\begin{equation}
     \int_1^\infty 
    \frac{u^{i\nu}\, \mathrm{d}u}{u\sqrt{u-1}}
    =
    B\left( \frac{1}{2}-i\nu, \frac{1}{2} \right)
    = \sqrt{\pi} \, 
    \frac{\Gamma \left( \frac{1}{2}-i\nu \right)}{\Gamma \left( 1-i\nu \right)} \, .
\end{equation}
Combining these results, we obtain the asymptotic expression given by Eq.~\eqref{eq:Kinu_result_asym}.

%


\end{document}